# Advancing Web Accessibility: A Guide to Transitioning Design Systems from WCAG 2.0 to WCAG 2.1

Hardik Shah

Department of Information Technology, Rochester Institute of Technology, Rochester, New York, USA

## Abstract

*This research focuses on the critical process of upgrading a Design System from Web Content Accessibility Guidelines (WCAG) 2.0 to WCAG 2.1, which is an essential step in enhancing web accessibility. It emphasizes the importance of staying up to date on increasing accessibility requirements, as well as the critical function of Design Systems in supporting inclusion in digital environments.*

*The article lays out a complete strategy for meeting WCAG 2.1 compliance. Assessment, strategic planning, implementation, and testing are all part of this strategy. The need for collaboration and user involvement is emphasized as critical strategies and best practices for a successful migration journey.*

*In addition, the article digs into migration barriers and discusses significant lessons acquired, offering a realistic view of the intricacies of this transforming road. Finally, it is a practical guide and a necessary resource for organizations committed to accessible and user-centered design. The document provides them with the knowledge and resources they need to navigate the changing world of web accessibility properly.*

## 1. Introduction

The WCAG (Web Content Accessibility Guidelines), established by the Web Accessibility Initiative (WAI) group under World Wide Web Consortium (W3C), serves as a globally recognized collection of guidelines and principles designed to make web content accessible to those with disabilities. This set of guidelines provides a structured approach to enhance the inclusivity and usability of digital content, including websites and web applications, for individuals with various disabilities, including visual, auditory, motor, and cognitive challenges. Web accessibility standards have become essential to government website compliance because they accord with equitable access, inclusion, and government agencies' legal responsibility to serve all residents [1], [2]. WCAG 2.1 significantly improves web accessibility requirements, expanding on the foundation established by WCAG 2.0. A critical component of the new criteria is that they address the changing landscape of digital interactions, with a particular emphasis on mobile accessibility. In today's world, when mobile devices are omnipresent, WCAG 2.1 recognizes this trend and provides criteria explicitly geared toward mobile platforms [3]. A significant success requirement, for example, is ensuring that all functionality can be accessed via touch gestures recognizing the widespread use of touchscreens on smartphones and tablets. This is especially important in situations where individuals with motor disabilities rely primarily on touch-based interactions to navigate and interact with digital material.

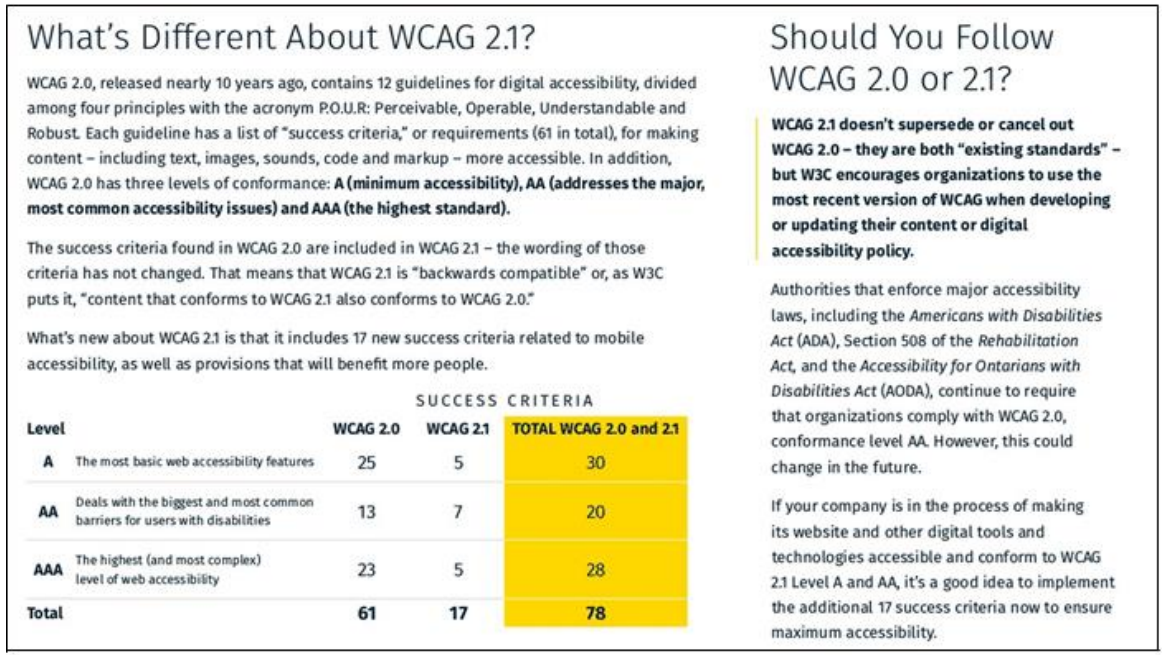

Figure 1: WCAG 2.1 scope and success criteria explained. Source: Adapted from [4]

Furthermore, WCAG 2.1 emphasizes the importance of delivering an inclusive experience for people with impaired vision [5]. New success criteria emphasize adjustable text spacing and contrast ratios in order to improve assistance for users with varied degrees of visual impairment. Consider the following scenario: a user with low vision visits a website on a desktop or mobile device [6]. The standards emphasize the necessity of ensuring that text is not just resizable but also adaptable in spacing, ensuring reading for people who need larger fonts or a unique visual presentation.

WCAG 2.1 also addresses cognitive and learning disabilities in depth. The new criteria emphasize developing a more cognitively accessible digital world, lowering possible barriers for people with various cognitive abilities. Consider the following scenario: a website with sophisticated terminology and extensive navigation [7]. The amended recommendations advocate for more straightforward language, predictable navigation, and fewer distractions, resulting in a more user-friendly experience for people with cognitive impairments [3]. Failure to comply with web accessibility regulations has resulted in litigation against governments and businesses [8]. Transitioning to WCAG 2.1 offers multiple benefits such as enhanced accessibility, a better user experience, adherence to legal standards, broader audience engagement, readiness for future developments, advantages in search engine optimization, ethical commitments, and a stronger position in the online marketplace [9]. It is a worthwhile investment promoting diversity while ensuring your digital information remains relevant and accessible in an ever-changing internet world. This research presents a clear roadmap for businesses and teams wanting to achieve WCAG 2.1 compliance by thoroughly examining the migration process. This roadmap includes assessing the accessibility status of the current Design System, understanding the subtleties of WCAG 2.1 standards, planning the migration, implementation, testing, and continuous compliance.

## 2. LITERATURE REVIEW

The evolution of web accessibility standards from the Web Content Accessibility Guidelines (WCAG) 2.0 to WCAG marks a significant shift in designing inclusive digital experiences [6]. This literature review explores various scholarly works and industry practices that guide the transition of design systems to comply with the updated standards.        Understanding WCAG 2.0 and its Limitations WCAG 2.0, established by the World Wide Web Consortium (W3C), has been the benchmark for web accessibility since its inception in 2008. It provided a comprehensive framework for making web content more accessible to people with disabilities [10]. However, research studies have pointed out its limitations, particularly in addressing the needs of users with cognitive disabilities and those relying on mobile devices. It emphasized the need for guidelines that evolve with technological advancements and also highlighted the gaps in WCAG 2.0 in catering to a broader range of disabilities [7].

### 2.1. The Emergence of WCAG 2.1

In response to these limitations, WCAG 2.1 was introduced in 2018. This version extends WCAG 2.0 by adding 17 additional success criteria focused on improving accessibility for mobile users, people with low vision, and those with cognitive and learning disabilities [8]. Another research work provides an in-depth analysis of these new WCAG 2.1 criteria, demonstrating how they enhance the user experience for a wider audience [9].

### 2.2. Transition Challenges and Strategies

The transition from WCAG 2.0 to 2.1 poses challenges for web developers and designers. They identified the need for updated training and awareness among professionals [11]. Similarly, a study conducted revealed the lack of preparedness in the industry for this transition, suggesting a need for comprehensive guidelines and tools to aid in the process [12].

### 2.3. Tools and Frameworks for WCAG 2.1 Compliance

Several researchers have developed tools and frameworks to assist in the transition. For instance, the work on the WAI-Tools Project provides automated testing tools that help in evaluating WCAG 2.1 compliance [13]. Additionally, a design framework has been introduced that integrates WCAG 2.1 principles into the design process, making accessibility a foundational component of web development [14].

### 2.4. Case Studies and Best Practices

Practical applications of WCAG 2.1 in real-world scenarios are crucial for understanding its impact. The study by [10] presents a case study of a university website's transition to WCAG 2.1, offering insights into best practices and challenges faced during the process [5]. Furthermore, [15] provided an analysis of how major corporations have adapted their design systems to comply with WCAG 2.1, highlighting the business benefits of accessibility.

## 2.5. Future Directions in Web Accessibility

Looking forward, studies like [16] discuss the future of web accessibility standards beyond WCAG 2.1. They emphasize the importance of continuous adaptation and the potential integration of emerging technologies like AI and machine learning in enhancing web accessibility.

In conclusion, the transition from WCAG 2.0 to WCAG 2.1 is a crucial step towards more inclusive web environments. The literature presents a comprehensive view of the challenges, strategies, and tools available for this transition. It also underscores the importance of ongoing research and development in the field of web accessibility to keep pace with technological advancements and the diverse needs of users.

## 3. IMPORTANCE OF DESIGN SYSTEMS

In the realm of web development, Design Systems have emerged as a fundamental framework, providing a structured approach to creating and managing digital products. These systems are not merely a collection of UI components and style guides; they represent a cohesive set of principles, patterns, and practices that guide the design and developmentprocess [6]. The importance of Design Systems lies in their ability to ensure consistency, improve efficiency, and foster collaboration among teams, ultimately leading to a more coherent user experience across various digital platforms.

A well-implemented Design System serves as a single source of truth for both designers and developers. It streamlines the design process by providing a library of reusable components and patterns [17]. This not only accelerates the development cycle but also ensures that the final product maintains visual and functional consistency [3]. By standardizing UI components, Design Systems reduce redundancy in the design process, allowing teams to focus on solving unique user problems rather than reinventing the wheel with each project. Moreover, Design Systems play a crucial role in enhancing the scalability of digital products [5]. As organizations grow and evolve, their digital products need to adapt without losing their core identity. Design Systems provide a flexible yet consistent framework that can accommodate new features and functionalities while maintaining the brand's visual language and user experience standards.

The strategic incorporation of Web Content Accessibility Guidelines (WCAG) 2.1 into Design Systems presents a significant advantage. WCAG 2.1 extends beyond the provisions of WCAG 2.0 by addressing a wider range of disabilities, including those related to vision, hearing, physical, speech, cognitive, language, learning, and neurological disabilities [16]. Integrating WCAG 2.1 directly into a Design System, as opposed to retrofitting accessibility into individual web applications, ensures that accessibility is not an afterthought but a foundational aspect of the design process.

This proactive approach to accessibility has several benefits. Firstly, it ensures that all components in the Design System are accessible from the outset, reducing the need for costly and time-consuming modifications later in the development process [10]. Secondly, it fosters an inclusive design philosophy, encouraging designers and developers to consider a diverse range of user needs and preferences from the beginning [12]. Through embedding WCAG 2.1 standards into the Design System, organizations can ensure compliance with legal requirements, thereby avoiding potential legal ramifications and enhancing their reputation as inclusive and socially responsible entities. In conclusion, Design Systems are indispensable in modern web development, offering a structured, efficient, and scalable approach to design and development.

The integration of WCAG 2.1 into these systems is not just a strategic advantage but a necessity in today's digital landscape, where accessibility and inclusivity are paramount [11]. By embracing this approach, organizations can create digital experiences that are not only aesthetically pleasing and consistent but also accessible to a broader audience, including those with disabilities.

## 4. UNDERSTANDING WCAG GUIDELINES AND KEY CHANGES IN WCAG 2.1

The Web Content Accessibility Guidelines (WCAG), established by the World Wide Web Consortium (W3C), serves as a globally recognized set of guidelines and principles designed to make web content accessible to those with disabilities. This set of guidelines provides a structured approach to enhance the inclusivity and usability of digital content, including websites and web applications, for individuals with various disabilities, including visual, auditory, motor, and cognitive challenges. Web accessibility standards have become essential to government website compliance because they accord with equitable access, inclusion, and government agencies' legal responsibility to serve all residents [1], [2]. Failure to comply with web accessibility regulations has resulted in litigation against governments and businesses [8]. WCAG 2.1 encompasses a mix of normative and informative guidelines [11], mirroring the structure found in WCAG 2.0, and it introduces an additional 17 success criteria aimed at advancing web accessibility [12]. Transitioning to WCAG 2.1 offers numerous benefits such as enhanced accessibility, an enriched user experience, adherence to legal standards, broader audience engagement, preparation for future requirements, advantages in search engine optimization, ethical considerations, and gaining a competitive edge in the online marketplace [9]. It is a worthwhile investment promoting diversity while ensuring your digital information remains relevant and accessible in an ever-changing internet world. This research presents a clear roadmap for businesses and teams wanting to achieve WCAG 2.1 compliance by thoroughly examining the migration process. The roadmap includes assessing the accessibility status of the current Design System, understanding the subtleties of WCAG 2.1 standards, planning the migration, implementation, testing, and continuous compliance.

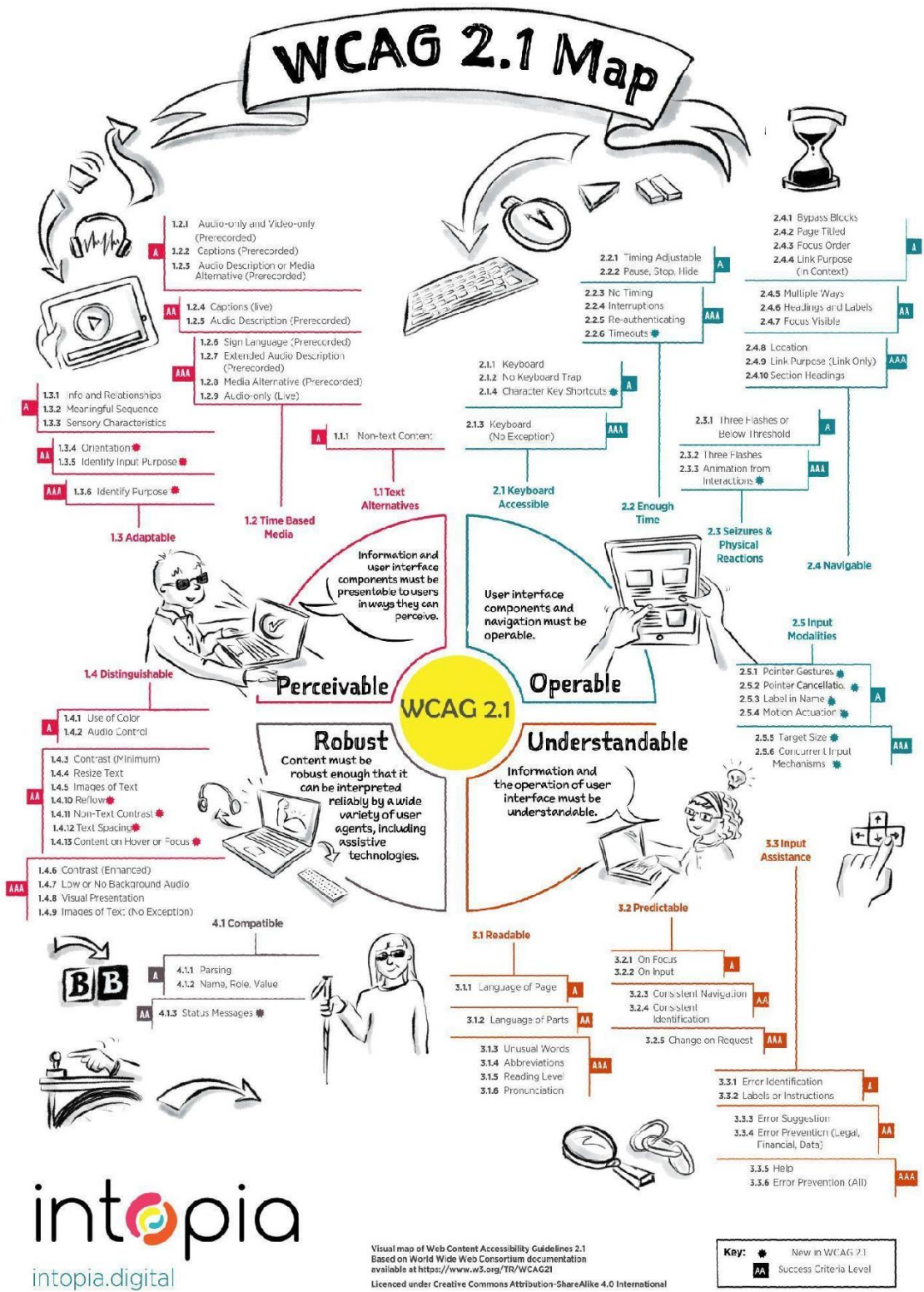

Figure 2: Key changes in WCAG 2.1. Source: Adapted from [13]

## 5.  BENEFITS OF ADDING WCAG 2.1 SUPPORT IN DESIGN SYSTEMS

In the evolving landscape of web development, the importance of accessibility cannot be overstated. The Web Content Accessibility Guidelines (WCAG) 2.1 represent a significant step forward in making web content more accessible to a wider range of people with disabilities [18]. Transitioning design systems from WCAG 2.0 to WCAG 2.1 is not just a compliance measure, but a strategic enhancement that brings numerous benefits.

### 5.1. Enhanced User Experience for a Broader Audience

WCAG 2.1 extends the accessibility considerations of WCAG 2.0 by including additional criteria to cater to users with cognitive and learning disabilities, users with low vision, and users with disabilities on mobile devices [19]. By more usable and inclusive, thereby reaching a wider audience [20]. This inclusivity is not only a moral imperative but also expands the potential user base, which can be particularly beneficial for commercial websites.

### 5.2. Improved Compliance with Legal Standards

Many countries are adopting stricter regulations regarding web accessibility. By aligning design systems with WCAG 2.1, organizations can ensure they are compliant with current and future legal requirements [21]. This proactive approach can prevent potential legal challenges related to accessibility, which can be costly and damaging to an organization's reputation.

### 5.3. Enhanced SEO and Online Visibility

Search engines increasingly favor websites with higher accessibility standards. WCAG 2.1's focus on clarity, navigation, and responsiveness contributes to better SEO. Websites that adhere to these guidelines are likely to rank higher in search engine results, leading to increased visibility and traffic [6].

### 5.4. Future-Proofing Web Assets

WCAG 2.1 is designed with future technologies in mind, including mobile and emerging assistive technologies. By
forward-thinking approach ensures that web assets remain relevant and accessible as new technologies emerge.

### 5.5. Enhanced Brand Image and Corporate Social Responsibility

Implementing WCAG 2.1 demonstrates an organization's commitment to diversity, equity, and inclusion [6]. This can enhance the brand's image and reputation, showing potential customers and partners that the organization values accessibility and inclusivity.

### 5.6. Reduced Maintenance and Development Costs

In the long run, incorporating WCAG 2.1 into design systems can lead to reduced maintenance and development costs. Accessible design is often cleaner and more efficient, leading to faster load times and reduced bandwidth usage [6]. Additionally, accessible websites tend to be more robust and easier to maintain, with fewer compatibility issues across different browsers and devices.

In conclusion, the integration of WCAG 2.1 into design systems is not just about adhering to standards; it is a strategic decision that enhances user experience, ensures legal compliance, improves SEO, future-proofs web assets, boosts brand image, and can lead to cost savings. As the digital world becomes increasingly inclusive, the transition from WCAG 2.0 to WCAG 2.1 is a crucial step for any organization committed to providing equitable access to its digital content.

## 6. MIGRATING A DESIGN SYSTEM FROM WCAG 2.0 TO WCAG 2.1

Transitioning from WCAG 2.0 to WCAG 2.1 in your Design System guarantees accessibility for all users, encompassing individuals with disabilities. This migration process entails a systematic approach to aligning your design system with the most recent accessibility requirements. The first step is to assess your existing WCAG 2.0 compliance [11]. This entails thoroughly assessing your existing Design Systems to determine which components already meet WCAG 2.0 criteria and which areas need improvement. You will detect accessibility difficulties, semantic markup practices, and color contrast concerns through automated and manual testing across the Design System's components, template layouts, and demonstration examples [1]. Documenting these findings and developing a repair plan is critical to resolving WCAG 2.0 compliance issues.

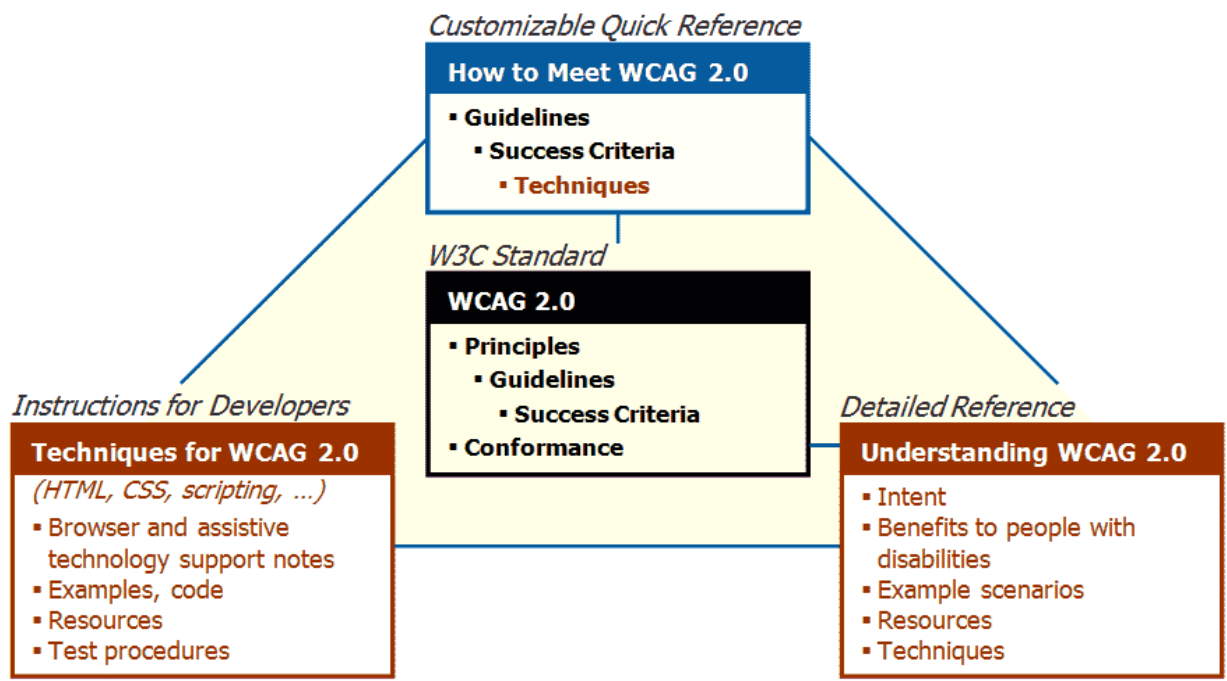

Figure 3: WCAG 2.0 migration - Quick Reference. Source: Adapted from [22]

When initiating the migration process, understanding the differences between WCAG 2.0 and WCAG 2.1 is essential. This information lays the groundwork for a smooth transition to the upgraded rules, ensuring you know the new success criteria, guidelines, and strategies provided in WCAG 2.1. The next step is to identify appropriate success criteria in WCAG 2.1 that are specific to your design system [21]. Because not all success criteria will have an immediate influence on your Design Systems, concentrate on those that will immediately impact the accessibility of your website or applications [14]. With a firm grasp of your starting point and the differences in the rules, it is essential to develop a complete migration strategy. This approach should encompass detailed methods, a timeline, designated roles and responsibilities for the migration process, and potential impacts on continuous design and development activities.

Several organizations have successfully transitioned from WCAG 2.0 to WCAG 2.1 by implementing effective tactics. Microsoft, for example, adopted a comprehensive effort to align its products, including Office 365 and Windows, with WCAG 2.1 criteria [21]. Microsoft promoted user interaction, doing extensive testing with people of varying abilities and incorporating comments to improve the accessibility of their products [19]. Another example is the BBC, which overhauled its design systems to meet WCAG 2.1 requirements. Prioritizing mobile accessibility and addressing cognitive disabilities through simplified language and navigation was part of the BBC's approach [5]. These examples highlight the significance of comprehensive assessment, strategic planning, and user participation in the migration process [6]. To ensure that their staff understood and supported the new standards, Microsoft and the BBC demonstrated proactive communication, transparent documentation, and regular training [21]. These real-world examples not only highlight successful transitions but also demonstrate the iterative nature of accessibility, urging firms to consider compliance as a journey rather than a one-time chore.

Prioritization is essential during your migration process. Prioritize accessibility improvements depending on their importance and urgency. Determine which situations require immediate attention and which can wait, allowing you to manage resources more effectively. As you progress, it is critical to review and update your accessibility rules and best practices documents [8]. Keeping your reference documents up to date can assist your design and development teams in keeping up with the current guidelines. The core of the migration process involves adopting the newly established success criteria set forth by WCAG 2.1 [11]. To achieve these revised accessibility standards, you must modify or add code, styles, and interaction patterns to your Design Systems [2]. The journey, however, continues after implementation. Conduct extensive accessibility testing to guarantee your Design Systems are entirely WCAG 2.1 compliant. This testing uses automated tools, manual evaluations, and assistive technology testing to identify and address any remaining issues.

Web accessibility is built on inclusivity; user testing and feedback are vital. Involve disabled persons in testing to obtain insights and feedback and fix any usability issues that may develop during this vital time. Provide training and awareness sessions on the new accessibility requirements set by WCAG 2.1 to empower your team for success [10]. It is critical for effective execution that your design and development teams grasp these requirements. Accessibility is a continuous effort. Create a procedure for continuing compliance that includes regular evaluations and changes to keep your Design Systems in line with WCAG 2.1 and future accessibility requirements. The importance of broad user participation in ensuring the success of the transition from WCAG 2.0 to WCAG 2.1 cannot be emphasized. Diverse user engagement gives a richness of viewpoints from a diverse range of skills, limitations, and user experiences [21]. This inclusivity serves as a litmus test for the efficacy of the modifications undertaken, allowing organizations to detect and address any accessibility hurdles that could otherwise go unnoticed [17]. Organizations obtain essential insights into the real-world usability of their digital assets by actively engaging users with varied needs, including those with visual, auditory, motor, and cognitive impairments. This user-centric approach not only adheres to essential accessibility standards but also develops a more empathetic and responsive design attitude.

Furthermore, incorporating varied users in the testing phase helps to create a digital environment that caters to a larger audience [21]. It guarantees that the improvements implemented not only meet compliance criteria but also resonate with end users, resulting in a genuinely inclusive online experience. As a result, the depth and diversity of user involvement throughout the testing process are inextricably related to the migration's success.

Throughout this procedure, effective communication is critical. Changes and upgrades to your Design System should be communicated as it transitions to WCAG 2.1 compliance, engaging stakeholders to ensure everyone is informed and on board with the accessibility advances [12]. Maintain thorough documentation of the migration process and seek expert support from accessibility experts or organizations as needed [15]. Their advice can be invaluable in ensuring full WCAG 2.1 compliance and offering an inclusive user experience. Finally, remember to see WCAG 2.1 compliance as a critical step toward establishing a more accessible and inclusive digital world [6]. It acknowledges the commitment and hard work of your team in enhancing the accessibility of digital content for a wider audience.

## 7. THE ROLE OF MANUAL AND AUTOMATED TESTING IN THE MIGRATION PROCESS

The significance of automated testing and manual audits in ensuring that Design Systems adhere to the most recent accessibility standards is critical when migrating from one set of accessibility guidelines to another, such as WCAG 2.0 to WCAG 2.1. Automated Scanning Tools are a helpful first step in the evaluation process. Organizations should proactively stay educated about developing rules as they anticipate future accessibility requirements such as WCAG 2.2 and WCAG 3.0 [6]. It is critical to embrace a culture of continual learning and adaptability [3]. Participating in pilot programs and early adoption activities for beta versions might provide valuable insights. Collaboration with user communities and the adoption of future technology, such as AI-powered accessibility solutions, will be critical [18]. Creating a structure for continuing accessibility reviews and encouraging a user-centric approach can help organizations move to and exceed forthcoming standards. This strategic vision not only assures compliance but also places enterprises at the forefront of providing inclusive digital experiences in a rapidly changing technology context. These tools are helpful for quickly finding common accessibility concerns in your Design System examples. Scanners excel at detecting flaws such as missing alt text for photos, incorrect markup structures, and text with insufficient contrast ratios [10]. They provide a rapid and systematic way to identify any issues with your Design System [5]. The W3C validation service, WebAIM Contrast Checker, Chrome Lighthouse, WAVE Web Accessibility Evaluation programs, and Accessibility Insights are well-known examples of automated scanning programs [14]. Utilizing these tools can accelerate the detection and resolution of simple difficulties. Manual testing, on the other hand, remains an essential component of the accessibility review process [1]. While automated technologies are beneficial, they may not detect all accessibility concerns. Human-led testing introduces a personal touch to the evaluation process, enabling a more detailed and subtle analysis [16]. It is critical to examine each design component in your Design System examples through the lens of accessibility during manual testing. This includes extensive testing with keyboard navigation, screen readers, and other assistive technology people with impairments use. Manual testing emphasizes key aspects such as proper management of focus, operability through keyboard input, and compatibility with screen readers. By doing extensive manual audits, you can uncover finer flaws that may not be detectable automatically.

## 8. CHALLENGES FACED IN WCAG 2.1 MIGRATION OF DESIGN SYSTEMS

During the transition from WCAG 2.0 to WCAG 2.1, organizations may face a number of problems that must be carefully considered. One significant source of concern is the possible resource strain, both in terms of time and labor, that will be necessary for the complete evaluation and implementation of new success criteria. Furthermore, updating existing digital assets to match the revised requirements may provide technological challenges, particularly for sophisticated systems or outdated applications [19]. The organizational culture's resistance to change may prevent the seamless adoption of WCAG 2.1 recommendations. Furthermore, ensuring that all team members have a thorough understanding of the new criteria may take time and effort [5]. Addressing these potential complaints necessitates a proactive approach that includes strong communication, resource planning, and ongoing training to promote a joint commitment to the long-term benefits of improved web accessibility [8]. Realistic expectations and a phased implementation strategy can reduce these issues, resulting in an easier transition and long-term compliance.

Organizations migrating from WCAG 2.0 to WCAG 2.1 may face a number of obstacles. One significant source of concern is the time and work required for a complete examination and implementation of new success criteria. Technological problems may occur, particularly for complicated systems or out-of-date applications, demanding extensive modifications to meet changed needs [17].

Furthermore, resistance to change within an organization's culture can inhibit the smooth implementation of WCAG2.1 recommendations. Another area for improvement is ensuring that all team members understand the complexities of the new criteria [7]. To address these difficulties, a proactive approach comprising excellent communication, resource planning, and continual training is required. Realistic expectations and a phased implementation strategy can allay fears, promoting a smoother transition and long-term compliance while respecting the nuanced challenges that companies may experience in their pursuit of improved web accessibility.

## 9. FUTURE WORK

The research on migrating Design Systems from WCAG 2.0 to WCAG 2.1 provides essential insights into the changing landscape of web accessibility [14]. However, certain limits must be acknowledged. The paper focuses primarily on technical factors, potentially ignoring the socio-cultural components of accessibility. Future research could address the intersectionality of accessibility, considering varied user requirements and experiences beyond the mentioned disabilities [6].

Additionally, while the handbook promotes collaboration, more significant inquiry into efficient interdepartmental cooperation and communication tactics during relocation could boost its practical application. The study's ramifications extend beyond compliance to broader ethical problems in digital design [20]. A more extensive investigation of the societal impact of accessible design and the potential for encouraging innovation and creativity in the digital world is a path for future inquiry.

## 10. CONCLUSION

Upgrading a Design System from Web Content Accessibility Guidelines (WCAG) 2.0 to WCAG 2.1 signifies a considerable commitment to encouraging diversity and user-centric design in the changing field of digital accessibility. As we conclude this study paper, it becomes evident that this transition represents more than merely a technological upgrade; it is a vital stride towards creating a digital environment that is both more accessible and equitable. The insights offered in this article emphasize that upgrading to WCAG 2.1 represents a transformative journey - one that closes the accessibility gap while also aligning with the larger objective of a universally inclusive digital world. It is critical to recognize that while this movement has its obstacles, it also opens up a world of potential. Furthermore, it positions companies and developers to effortlessly move to future accessibility standards like WCAG 2.2 and WCAG 3.0, ensuring that digital experiences continue to improve in a way that benefits all users.

## AUTHORS

**Hardik Shah**, completed his MS in IT degree from Rochester Institute of Technology, New York, U.S.A. and has 12+ years of professional experience with UX Design and full-stack Web Development. Currently, he specializes in leading Web UI development teams and building semantic and accessible Design Systems.

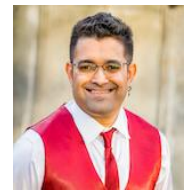